\documentstyle[12pt,rotate,cite]{article}
\newcommand{\be}{\begin{equation}}
\newcommand{\ee}{\end{equation}}
\newcommand{\bea}{\begin{eqnarray}}
\newcommand{\eea}{\end{eqnarray}}
\newcommand{\bd}{\begin{displaymath}}
\newcommand{\ed}{\end{displaymath}}
\newcommand{\f}{\frac}
\newcommand{\ra}{\rightarrow}

\newcommand{\bc}{B_c} 
\newcommand{\Bcet}{$B^-_c \ra \eta^{\prime} l^- \bar{\nu} $ }
\newcommand{\bcet}{B_c^- \ra \eta^{\prime} l^- \bar{\nu}  }

\newcommand{\gev}{\, {\rm GeV}}
\newcommand{\mev}{\, {\rm MeV}}
\begin{document}
\bibliographystyle{physics}
\renewcommand{\thefootnote}{\fnsymbol{footnote}}

\author{
Akio Sugamoto  and    Yadong Yang \thanks{JSPS Fellow}
\thanks{sugamoto@phys.ocha.ac.jp, yangyd@phys.ocha.ac.jp}\\
{\small\sl  Department of Physics, Ochanomizu  University,
1-2 Otsuka 2 Chome Bunkyo-ku, Tokyo 112, Japan }
}
\date{}
\title{
\vspace{3cm}
\bigskip
\bigskip
{\LARGE\sf The annihilation decay 
 \Bcet  }
}
\maketitle
\thispagestyle{empty}
\begin{abstract}
\noindent
We first investigate the semileptonic annihilation decay \Bcet in QCD.
We find the $\eta^{\prime}$ momentum distribution is peaked within its
small recoil region due to the loop effects. The branching ratio is
estimated to be $ Br(\bcet)=1.6\times 10^{-4}$ for $l=\mu,e$, which is
accessible at CERN LHC. 

\vspace{5cm}
PACS numbers 14.40 Nd 13.20 He 13.30 Ce 12.38.Bx 
\end{abstract}
\newpage
\setcounter{page}{1}

\setcounter{footnote}{0}
\renewcommand{\thefootnote}{\arabic{footnote}}
\section{Introduction}
Being consisted of two different heavy flavors, the bottom-charmed
meson $\bc$ have many fascinating properties, which have motivated 
extensive studies in the literature. Its productions\cite{production},
spectroscopy\cite{mass, quigg1} and
decays\cite{decay, beneke} could  be estimated to  certain accuracy and 
provid windows for probing both strong and weak interactions. 

The recent observation of $\bc$ in 1.8 TeV $p\bar p$ collisions using 
CDF detector at the Fermilab Tevatron has confirmed its existence in nature
with mass $M_{\bc}=6.40\pm0.39\pm0.13\gev$ and lifetime 
$\tau_{\bc}=0.46^{+0.18}_{-0.16}\pm0.03ps$, which agree with the
theoretical predictions\cite{production, beneke}. Further detailed
experimental studies will 
be performed at Tevatron Run II and CERN large Hadron Collider(LHC).
Especially, at LHC with the luminosity ${\cal L}=10^{34}cm^{-2}s^{-1}$ 
and $\sqrt{s}=14\rm TeV$, the number of $\bc^{\pm}$ events is expected 
to be about $10^{8}\sim10^{10}$ per year, so that some $\bc$ rare decays
of interests could be studied.

In this paper, we would like to present the first investigation on the
annihilation decays \Bcet  in  QCD. In this process, $b$ and $\bar c$
annihilate to leptons pair and meanwhile emit two gluons to form 
$\eta^{\prime}$. Due to OZI suppression, its decay width
should be two or three 
orders of magnitude lower than the dominant semileptonic decays induced by
$b\rightarrow c$
weak current. So it belongs to rare decays. However it is still sizable
at LHC.

Compared with the pure leptonic decays $\bc \ra l \bar{\nu}_l$, 
the suppression 
factor is  $\alpha_s^4 $ in $\Gamma$(\Bcet) instead of the helicity
suppression factor $\frac{m^2_l}{m_{B_c}^2}$. Considering the three bodies 
phase space much smaller than two bodies phase space, one could
expect that the decay width $\Gamma$(\Bcet) may be the same order 
as $\Gamma(\bc \ra \mu {\bar \nu}_{\mu})$ , anyhow it would be much 
larger than 
$\Gamma(\bc \ra e {\bar \nu}_{e})$. As we know from the radiative $J/\Psi$
decays, the coupling  $g^* g^* \ra \eta^{\prime}$ is much larger then 
$g^* g^* \ra \eta, \pi^0 $. For example, the ratio $BR(J/\Psi \ra \gamma
\eta^{\prime})/ BR(J/\Psi \ra \gamma \pi^0 )$ is as large as $10^2$. In
the following, the branching ratio is
found to be $Br(\bcet)\sim 10^{-4}$ for $l=\mu, e$. It is of interests
that 
the common start point suppression factor $\mid k_{\eta^{\prime}}\mid$
arising from 
the phase space integration $d^3 k_{\eta^{\prime}}$  is canceled by the
loop functions, so, the distribution
$dBr(B_{c} \ra \eta^{\prime}l{\bar\nu})/dE_{\eta^{\prime}}$ is peaked
within the small recoil region of $\eta^{\prime}$. This feather makes the
decay
recognizable from the mean experimental background $B_u^- \rightarrow
\eta^{\prime}l {\bar\nu}$. 

This paper is organized as followings. In section 2, we give the details
of the calculation of  the amplitude and phrase space integration.
Section 3 is devoted to numerical results and discussions.

\section{Calculations}

To order of $\alpha^4_s$, the process \Bcet is described by 6 diagrams
as    in Fig.1. It is easy to understood that  the heavy
quarkonium like $B_c$ bound state justifies perturbative QCD calculations
of the  decay. The method outlined years ago for $J/\Psi \rightarrow
\eta^{\prime}\gamma$ decays by $Korner, et al.,$\cite{korner} is
appropriate for the
present case. However, we would like to adopt an effective
Lagrangian approach to
avoid introducing $B_c$ meson wave function. At first, we begin with
the sub-amplitude of $(b\bar{c})\rightarrow g^*_a g^*_b l\bar\nu$. 
\bea
{\cal M} (b\bar c \ra g^*_a g^*_b l^- \nu )=
\f{G_{F}}{\sqrt{2}}
 V_{cb}g_s^2 Tr[T_a T_b] \bar{v}_c (p_c ) 
 \left[ 
       \gamma_{\mu}(1-\gamma_{5} )
  \frac{i}{\slash{\hskip -0.23cm}p_{b} -\slash{\hskip -0.23cm}K -m_b}
  \gamma_{\beta}
  \frac{i}{\slash{\hskip -0.23cm}p_{b} -\slash{\hskip -0.23cm}k_1 -m_b}
  \gamma_{\alpha} \right. \nonumber\\
\left.
+\gamma_{\alpha}
\frac{i}{\slash{\hskip -0.23cm}k_{1} -\slash{\hskip -0.23cm}p_{c}-m_c}
  \gamma_{\beta}
\frac{i}{\slash{\hskip -0.23cm}K-\slash{\hskip -0.23cm}p_{c}-m_c}
  \gamma_{\mu}(1-\gamma_{5} )
+\gamma_{\beta}
\frac{i}{-\slash{\hskip -0.23cm}p_{c}+\slash{\hskip -0.23cm}k_{2} -m_c}
  \gamma_{\mu}(1-\gamma_{5} )
  \frac{i}{\slash{\hskip -0.23cm}p_{b}-\slash{\hskip -0.23cm}k_{1} -m_b}
  \gamma_{\alpha}\right] u_{b}(p_{b})
\nonumber \\
\times
  {\bar l}\gamma^{\mu}(1-\gamma_{5})\nu_{l} 
+\left(\alpha\leftrightarrow\beta, k_{1}\leftrightarrow k_{2} \right).
\eea
Using the identity
\be
\gamma^{\mu}\gamma^{\alpha}\gamma^{\beta}=\gamma^{\mu}g^{\alpha\beta}
+\gamma^{\beta}g^{\mu\alpha}-\gamma^{\alpha}g^{\mu\beta}
-i\epsilon^{\mu\alpha\beta\delta}\gamma_{\delta}\gamma_{5}
\ee
and  Dirac equation $(\slash{\hskip -0.23cm}p -m)u(p)=0$, after a bit of
algebra, we arrive at an effective Lagrangian which has the form
\be
{\cal A}=
\frac{G_{F}}{\sqrt{2}}
 V_{cb}g_{s}^{2} Tr[T_{a}T_{b}]
 \bar{c}\gamma_{\delta}(1-\gamma_{5})
b{\hskip 2mm} {\bar l}\gamma_{\mu}(1-\gamma_{5})\nu_{l}
 {\cal F}^{\delta\mu\alpha\beta}\f{1}{k^2_1}\f{1}{k^2_2}
\langle g^*_{a\alpha} g^*_{b\beta} \mid\eta^{\prime}\rangle.
\ee
Then we can use the definition 
\be
\langle 0\mid \bar{c}\gamma_{\mu}(1-\gamma_{5})b\mid B_{c}(P)\rangle=
if_{B_c}P_{\mu}
\ee
and the $g^*_a g^*_b \ra\eta^{\prime}$ coupling 
\be
\langle g^*_a g^*_b \mid\eta^{\prime}\rangle=g_s^2 \delta_{ab}
\frac{A_{\eta^{\prime}} }{k_{1}\cdot k_{2}}\epsilon_{\alpha\beta m n }
k_{1}^m k_{2}^n
\ee
which has been  widely used in $\eta^{\prime}$ and pseudoscalar productions
in heavy quarkonium decays and in high energy collidors\cite{close}. Here
the parameter
$A_{\eta^{\prime}}$ is understood as a combination of $SU(3)$ mixing
angles and nonperturbative
objects,  and can be extracted from the decay
$J/\Psi\ra\eta^{\prime}\gamma$.
We obtain the total amplitude as
\bea
{\cal M}&=&\frac{G_{F}}{\sqrt{2}}
 V_{cb}g_s^4 Tr[T_a T_b]\delta_{ab}4A_{\eta^{\prime}} if_{B_c} 
 \bar{l}\gamma^{\mu}(1-\gamma_{5})\nu_{l}   \nonumber\\
& &\frac{1}{2} \left[ 
\frac{
2P_{\mu}k_{\delta}+2P_{\delta}K_{\mu}-2P{\cdot}K g_{\mu\delta}
+2i\epsilon_{\mu\rho\delta\sigma}P^{\sigma}K^{\rho}
+4M_{B_c}m_{b}g_{\mu\delta}-4p_{b\mu}P_{\delta}
}
{K^2 -2K{\cdot}p_{b}} \right. \nonumber\\
& &\ \times
\int\frac{d^{4}q}{(2\pi)^4}
\left[
\frac{1}{k_{1}^2 -2p_{b}{\cdot}k_{1}}
\left( 
     \frac{ k_{1}^{\delta} }{ k_{1}^{2}k_{2}^{2} } 
    -\frac{ k_{2}^{\delta} }{ k_{2}^{2}k_{1}{\cdot}k_{2} }
\right)
-\frac{1}{k_{2}^2 -2p_{b}\cdot k_{2}}
\left( 
     \frac{ k_{1}^{\delta} }{ k_{1}^{2}k_{1}{\cdot}k_{2} } 
    -\frac{ k_{2}^{\delta} }{ k_{1}^{2}k_{1}^{2} }
\right)
\right]   \nonumber\\
& &+
\frac{
2P_{\mu}k_{\delta}+2P_{\delta}K_{\mu}-2P{\cdot}K g_{\mu\delta}
+2i\epsilon_{\mu\rho\delta\sigma}P^{\sigma}K^{\rho}
+4M_{B_c}m_{c}g_{\mu\delta}-4p_{c\mu}P_{\delta}
}
{K^2 -2K{\cdot}p_{c}}  \nonumber\\
& & \times
\int\frac{d^{4}q}{(2\pi)^4}
\left[
\frac{-1}{k_{1}^2 -2p_{c}{\cdot} k_{1}}
\left( 
     \frac{ k_{1}^{\delta} }{ k_{1}^{2}k_{2}^{2} } 
    -\frac{ k_{2}^{\delta} }{ k_{2}^{2}k_{1}{\cdot}k_{2} }
\right)
-\frac{1}{k_{2}^2 -2p_{c}{\cdot}k_{2}}
\left( 
     \frac{ k_{1}^{\delta} }{ k_{1}^{2}k_{1}{\cdot}k_{2} } 
    -\frac{ k_{2}^{\delta} }{ k_{1}^{2}k_{1}^{2} }
\right)
\right]  \nonumber \\
& & 
+\int\frac{d^{4}q}{(2\pi)^4}
 \frac{2(k_{2\mu}k_{1}^{2}P\cdot k_{2}-k_{1\mu}k_{2}^{2}P{\cdot}k_{1}}
 {k_{1}^{2}k_{2}^{2}k_{1}{\cdot}k_{2}}
\left. \left(
\frac{1}{ k_{1}^{2}-2k_{1}{\cdot}p_{b}}
\right) \right],
\eea
where $K=k_{1}+k_{2}$ is the momentum of $\eta^{\prime}$, $P$ is the
monetum of $B_c$  and $q$ is the loop
momentum with the relation $2q=k_1 -k_2$, 
and the factor
$\frac{1}{2}$ takes into account that both sub-amplitudes have already
been symmetrized with respect to the two gluons.

For the heavy $b$ and $c$ quarks, it is reasonable to neglect the
relative momentum of the quark constituents and their binding energy
relative to their masses. In this nonrelativistic limit, the
constituents are on mass shell and move together with the same velocity.
It implies the following equations valid to good accuracy
\be
M(B_{c})=m_{c}+m_{b}, {\hskip 0.5cm} p_{\bar c}=\frac{m_c}{M}P,
{\hskip 0.5cm}p_{b}=\frac{m_b}{M}P.
\ee
Hence, we have omitted the terms in eq(6) which are  proportional to
\be
\epsilon^{\mu\nu\alpha\beta}p_{b\mu}p_{{\bar c}\nu}\cdots.
\ee

It is straightforward to perform the loop integration in eq(6) using
Dimensional Regularization. The amplitude is found to be 
\be
{\cal M}=\frac{G_{F}}{\sqrt{2}}
 V_{cb}g_s^4 Tr[T_a T_b]\delta_{ab}4A_{\eta^{\prime}} if_{B_c} 
 \frac{i}{16\pi^2}\left( P_{\mu}f_{1}+K_{\mu}f_{2}\right)
 \bar{l}\gamma^{\mu}(1-\gamma_{5})\nu_{l}
\ee
with $f_{1}, f_{2}$ defined by
\bea
f_{1}&=&-4C_{11}(K, p_{b}-K, 0, 0, m_{b})+4C_{12}(K, p_{b}-K, 0, 0,m_{b})
\nonumber \\
&&-2C_{11}(\frac{K}{2}, \frac{K}{2}-p_{b}, 0,
       \frac{m_{\eta^{\prime}}}{2}, m_{b})
   -2C_{12}(\frac{K}{2}, \frac{K}{2}-p_{b}, 0,
      \frac{m_{\eta^{\prime}}}{2}, m_{b})
\nonumber \\
&&-4C_{11}(K, p_{c}-K, 0, 0, m_{c})+4C_{12}(K, p_{c}-K, 0, 0,m_{c})
\nonumber \\
&&+2C_{11}(\frac{K}{2}, \frac{K}{2}-p_{c}, 0,
       \frac{m_{\eta^{\prime}}}{2}, m_{c})
   +2C_{12}(\frac{K}{2}, \frac{K}{2}-p_{c}, 0,
      \frac{m_{\eta^{\prime}}}{2}, m_{c})
\nonumber \\
&& +\frac{2m_{b}}{m_{c} }C_{12}(\frac{K}{2}, p_{b}-K, 0,
       \frac{m_{\eta^{\prime}}}{2}, m_{b})
-\frac{2m_{c}}{m_{b} }C_{12}(\frac{K}{2}, p_{c}-K, 0,
       \frac{m_{\eta^{\prime}}}{2}, m_{c})
\nonumber \\
&&
-\frac{2M(m_{b}-m_{c})}{m_{b}m_{c}}\left( 
C_{12}(\frac{K}{2}-p_{c}, P-K,
       \frac{m_{\eta^{\prime}}}{2}, m_{c}, m_{b})
\right.  \\
&& \left.
-\frac{m_{c}}{M}C_{11}(\frac{K}{2}- p_{c}, P-K, 0,
       \frac{m_{\eta^{\prime}}}{2}, m_{c}, m_{b}) \right), \nonumber
\eea
and
{\small
\bea
&f_{2}&=\frac{-4Mm_{b}}{K^2 -2p_{b}{\cdot}K} 
\left(
2C_{11}(K, p_{b}-K, 0, 0, m_{b})-C_{12}(K, p_{b}-K, 0, 0,m_{b})
+C_{11}(\frac{K}{2}, \frac{K}{2}-p_{b}, 0,
       \frac{m_{\eta^{\prime}}}{2}, m_{b})
\right)  \nonumber \\
&&+\frac{4Mm_{c}}{K^2 -2p_{c}{\cdot}K}
\left(
2C_{11}(K, p_{c}-K, 0, 0, m_{c})-C_{12}(K, p_{c}-K, 0, 0,m_{c})
+C_{11}(\frac{K}{2}, \frac{K}{2}-p_{c}, 0,
       \frac{m_{\eta^{\prime}}}{2}, m_{c})
\right)
\nonumber \\
&&+\frac{M}{m_{c}}
\left( 
C_{11}(\frac{K}{2}, p_{b}-K,
       \frac{m_{\eta^{\prime}}}{2}, 0, m_{b})
-2C_{12}(\frac{K}{2}, p_{b}-K,
       \frac{m_{\eta^{\prime}}}{2}, 0, m_{b})
+C_{0}(\frac{K}{2}, p_{b}-K,
       \frac{m_{\eta^{\prime}}}{2}, 0, m_{b}) 
\right)
\nonumber \\
&&-\frac{M}{m_{b}}
\left( 
C_{11}(\frac{K}{2}, p_{c}-K,
       \frac{m_{\eta^{\prime}}}{2}, 0, m_{c})
-2C_{12}(\frac{K}{2}, p_{c}-K,
       \frac{m_{\eta^{\prime}}}{2}, 0, m_{c})
+C_{0}(\frac{K}{2}, p_{c}-K,
       \frac{m_{\eta^{\prime}}}{2}, 0, m_{c})
\right)
\nonumber \\
&&-\frac{M(m_{b}-m_{c})}{m_{b}m_{c}}
\left(
C_{11}(\frac{K}{2}-p_{c}, P-K,
       \frac{m_{\eta^{\prime}}}{2}, m_{c}, m_{b})
-2C_{12}(\frac{K}{2}-p_{c}, P-K,
       \frac{m_{\eta^{\prime}}}{2}, m_{c}, m_{b})       
\right. \nonumber \\
&&\left. +C_{0}(\frac{K}{2}-p_{c}, P-K,
       \frac{m_{\eta^{\prime}}}{2}, m_{c}, m_{b})
\right)
\eea
}
The scalar loop functions and their definitions can be found in 
ref\cite{veltman}.
The divergences are canceled as they should be. 

With eq(9), we get
\bea
\frac{dBr(\bcet)}{dE_{\eta^{\prime}}}
=\frac{1}{(2\pi)^5}\frac{1}{16M_{B_c}}4\pi
\sqrt{E^{2}_{\eta^{\prime}}-m^{2}_{\eta^{\prime}}}
\frac{16\pi}{3}M_{B_c}^2 (E^{2}_{\eta^{\prime}}-m^{2}_{\eta^{\prime}} )
\mid f_1 +f_2 \mid^2 C^2 \tau_{B_c},
\eea
where
\be
C=\frac{8}{3}\alpha^{2}_s f_{B_c}A_{\eta^{\prime}}
\frac{G_F}{\sqrt{2}}V_{cb}.
\ee
\section{Numerical Results and Discussions}
For numerical results, we would take $\alpha_s =\alpha_s (M_{\bc})=0.2$,
 $V_{cb}=0.04$, $A_{\eta^{\prime}}=0.2$ and $\tau_{B_c} =0.46ps$. 
The decay constant $f_{\bc}$ probes the strong(nonpertubative) QCD 
dynamics which bind $b$ and $\bar c$  quarks to form the bound state $\bc$.
It is common wisdom to realize that the size of $\bc$ would  much 
larger than the size of $B_{\bar q}$. The size of $\bc$  scales as
$1/m_c$, but the size of $B_{\bar q}$  scales as $1/m_{\bar q}$ 
$(q=u, d, s)$. The compact size of $\bc$ would enhance the importance of 
its annihilation decays and imply the decay constant $f_{\bc}$ would 
much larger than $f_{B}$\cite{quigg1}.  In nonrelativistic limit, 
$f_{\bc}$ can be related to the value of its wave function at origin\cite{van}.
Using the nonrelativistic potemtal models, Echiten and Quigg \cite{quigg1}
 estimated 
\be
f_{\bc}=\left\{\begin{array}{ll}
       500\mev & \mbox{(Buchm{\"u}ller-Tye potential\cite{p1})}\\
       512\mev & \mbox{(power law potential\cite{p2})}\\
       479\mev & \mbox{(logarithmic potential\cite{p3})}\\
       687\mev & \mbox{(cornell potential\cite{p4})}
\end{array}\right.  
\ee
For numerical illustrations, we would take 
$f_{\bc}=500\mev$.
The $\eta^{\prime}$ momentum distribution is displayed in Fig.2. We find the
$\eta^{\prime}$ momentum distribution is peaked within small recoiling
region of $\eta^{\prime}$. One can expand the scalar functions in $f_1$
and $f_2$
in terms of basic scalar functions $B_0$ and $C_0$ in \cite{veltman} and
find the factor
$\sqrt{E^{2}_{\eta^{\prime}}-m^2_{\eta^{\prime}}}
(E^{2}_{\eta^{\prime}}-m^2_{\eta^{\prime}})$
will be canceled by the loop function. For an example, we expand
$C_{11}(K, p_{b}-K, 0, 0, m_{b})$ as
\bea
C_{11}(K, p_{b}-K, 0, 0, m_{b})                
=\f{1}{2(K^{2}(p_b -K)^2 -(K\cdot(p_b -K))^2 )}{\hskip 5cm}\\ \nonumber
\times\left[(p_b -K)^2 ( B_{0}(p_{b},0,m_b )-B_{0}(p_{b}-K,0,0 )
-K^2 C_0 (K, p_{b}-K, 0, 0, m_{b}))+\cdots\right]
\\ \nonumber
=-\frac{1}{2m_b^2 (E^{2}_{\eta^{\prime}}-m^2_{\eta^{\prime}}) }
\times\left[\cdots\right].{\hskip 6.5cm} 
\eea
Therefore, the $\eta^{\prime}$ momentum distribution would behave as
\be
{\propto}\frac{1}{ \sqrt{E^{2}_{\eta^{\prime}}-m^2_{\eta^{\prime}}
}},
\ee
when $E_{\eta^{\prime}}$ is small. The singularity at the start point 
of the  distribution due  to  the factor in eq(15) is integratable and 
give  finite decay width. Such peculiar property would make the decay
itself   recognizable from its mean background 
$B_u^- \ra \eta^{\prime}l\bar\nu$
at LHC, especially, when the decay chain $\eta^{\prime}\ra \gamma \gamma$
could be used to reconstruct the events in data analysis. 

The branching ratio is estimated to be 
\be
Br(\bcet)=1.6\times 10^{-4},
\ee
which is accessible at LHC. We can extend the estimation to 
$Br(B_c \ra \pi^0 l \bar\nu)$, if the following relation is valid 
\bea
\frac{Br(B_c \ra \eta^{\prime}\ell\bar\nu)}{Br(B_c \ra \pi^0 \ell \bar\nu)}
{\simeq} 
\frac{Br(J/\Psi \ra \eta^{\prime}\gamma )}{Br(J/\Psi \ra \pi^{0}\gamma)}   
{\simeq} 10^2 ,
\eea
which is the ratio square of the relative $\eta^{\prime}$ and $\pi^0$ 
coupling strengths to gluons, when the phase space diference 
is small due to $m_{\eta^{\prime}}$ and $m_{\pi^0}$. In this way, we get 
\be
Br(B_c \ra \pi^0 \ell\bar\nu)\simeq 1.5\times 10^{-6}.
\ee

In conclusion, we have presented the first study on the semileptonic 
annihilation decays \Bcet. The $\eta^{\prime}$ momentum distribution 
in the decay is found peaking within the small recoil region of
$\eta^{\prime}$ because of the loop effects, which is different from
the common tree level cases. The branching ratio is estimated to be 
few times larger than the pure leptonic decays $B_c^- \ra \mu
\bar\nu$\cite{yang} and $10^{4}$ times larger than $B_c^- \ra e\bar\nu$. With
large samples to be obtained at LHC, the
decay could be measured  and might be used to extract $B_c$ decay constant
and/or to probe strong and weak interactions. 
      
%%%%%%%%%%%%%%%%%%%%%%%%%%%%%%%%%%%%%%%%%%%%%%%%%%%%%%%%%%%%%%%%%%%%%%%%%
% Acknowledgment and bibliography
%%%%%%%%%%%%%%%%%%%%%%%%%%%%%%%%%%%%%%%%%%%%%%%%%%%%%%%%%%%%%%%%%%%%%%%%%
\bigskip
\noindent
{\large\bf Acknowledgment}

\noindent
One of us (Y. Yang ) would like to thank the support of
Japan Society for Promotion of Science(JSPS). This work is also supported 
in part by the Grant-in-Aid for Scientific Research No.08640357,
No.08640400 and the Grant-in-Aid for Research on Priority Areas( Physics
of CP Violation, No.10140208) from the Ministry of Education, Science and
Culture, Japan.

\newpage
\begin{center}
\large{Figure Captions}
\end{center}
\vspace{4cm}
Figure 1: Diagrams for \Bcet at the leading level. The blob represents  
$\eta^{\prime}$\\
\\
\\
\\
\\
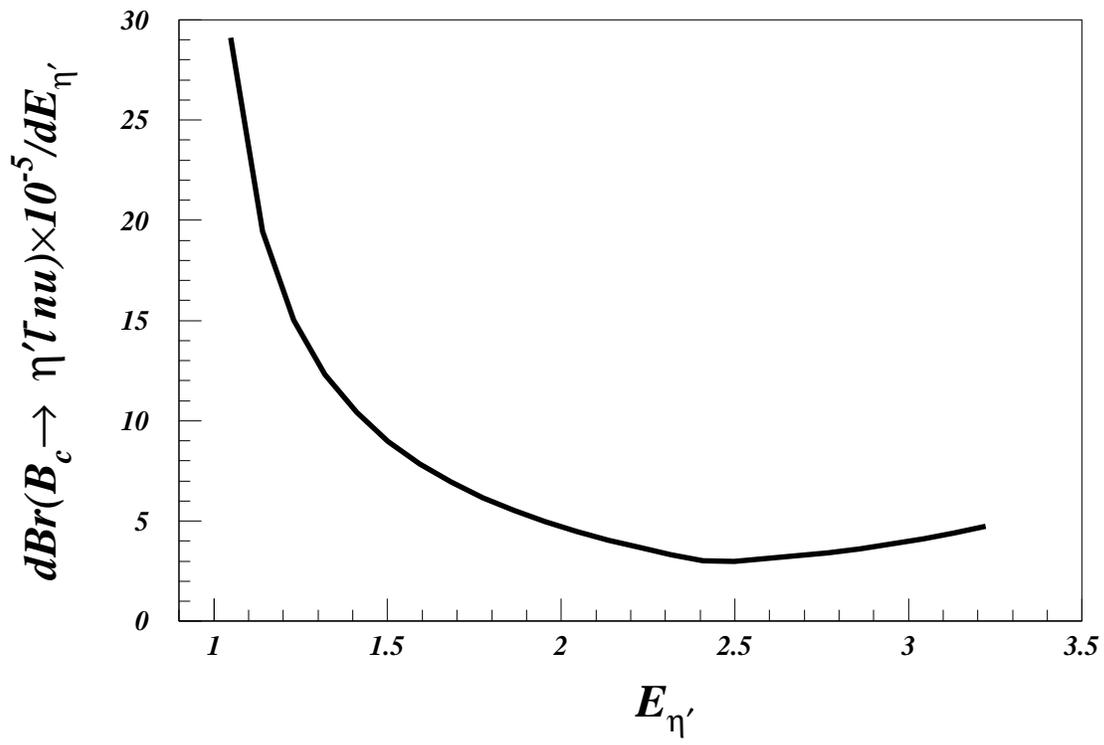
Figure 2: The distribution of $dBr(\bcet)/dE_{\eta^{\prime}}$ as a
function of $E_{\eta^{\prime}}$

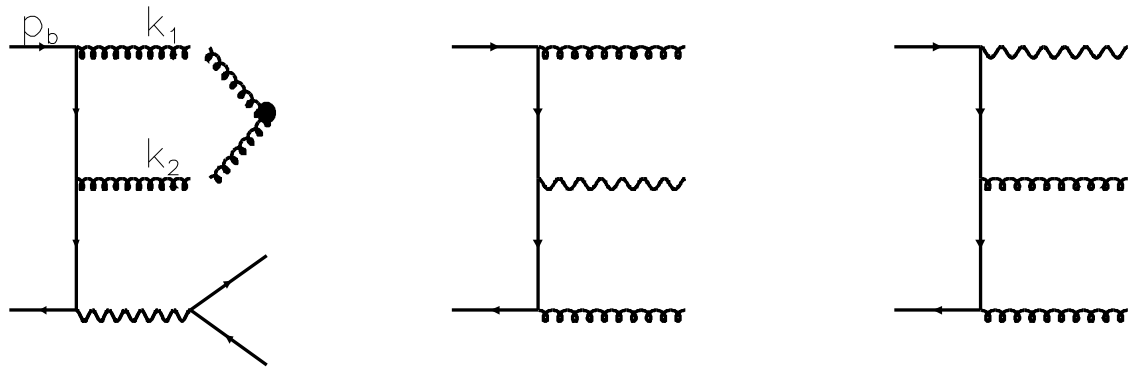
\end{document}